\documentclass[11pt,a4paper]{article}
\pdfoutput=1
\usepackage[T1]{fontenc}
\usepackage{jcappub}

\newcommand{\be}{\begin{equation}}
\newcommand{\ee}{\end{equation}}
\newcommand{\bea}{\begin{eqnarray}}
\newcommand{\ena}{\end{eqnarray}}
\newcommand{\no}{\noindent}

\newcommand\m{\ensuremath{\mu}}

\newcommand{\de}{\partial}

\newcommand{\ba}{\begin{eqnarray}}
\newcommand{\ea}{\end{eqnarray}}

\title{\boldmath $\Delta N$ Formalism and Conserved Currents in Cosmology}

\author[a,b,c,d]{Sabino Matarrese,}
\author[e,f]{Luigi Pilo}
\author[d]{and Rocco Rollo}

\affiliation[a]{Dipartimento di Fisica e Astronomia ``G. Galilei",
Universit\`{a} degli Studi di Padova,\\
 via Marzolo 8, I-35131 Padova,Italy}
\affiliation[b]{INFN, Sezione di Padova,\\
 via Marzolo 8, I-35131 Padova, Italy}
\affiliation[c]{INAF-Osservatorio Astronomico di Padova,\\
Vicolo dell' Osservatorio 5, I-35122 Padova, Italy}
\affiliation[d]{Gran Sasso Science Institute (GSSI),\\
Viale Francesco Crispi 7, I-67100 L'Aquila, Italy}
\affiliation[e]{Dipartimento di Ingegneria e Scienze dell'Informazione
  e Matematica, Universit\`a degli Studi dell'Aquila,\\
Via Vetoio (Coppito 1),I-67010 L'Aquila, Italy}
\affiliation[f]{INFN, Laboratori Nazionali del Gran Sasso,\\
I-67010 Assergi, Italy}

\emailAdd{rocco.rollo@gssi.it}
\emailAdd{sabino.matarrese@pd.infn.it}
\emailAdd{luigi.pilo@aquila.infn.it}

\abstract{The $\Delta N$ formalism,  based on
the counting of the number of e-folds during inflation in different
local patches of the Universe, has been introduced several years ago as a simple and physically intuitive approach to calculate (non-linear) curvature 
perturbations from inflation on large sales, without resorting to the
full machinery of (higher-order) perturbation theory. Later on, it was
claimed the equivalence with the results found by introducing a conserved fully non-linear
current $\zeta_\mu$, thereby allowing to directly connect
perturbations during inflation to late-Universe observables. We
discus some issues arising from the choice of the initial
hyper-surface in the  $\Delta N$ formalism.
By using a novel exact expression for $\zeta_\mu$, valid for any barotropic
fluid, we find that it is not in general related to the standard uniform
density curvature perturbation $\zeta$; such a result conflicts with the
claimed equivalence with  $\Delta N$ formalism.
Moreover, a similar analysis is done for
the proposed non-perturbative generalization ${\cal R}_\mu$ of the comoving 
curvature perturbation ${\cal R}$.}


\arxivnumber{1812.03844}

\begin{document}
\maketitle
\flushbottom
\section{Introduction}
\label{in_sec}
Thanks to technologies development and satellite missions such us WMAP and {\it Planck}, an unprecedented level of precision ~\cite{Aghanim:2018eyx} has been achieved in measurements of Cosmic Microwave Background (CMB) temperature anisotropies and polarization. The main properties of these fluctuations are well explained by the inflationary paradigm; however a strong model degeneracy persists, which  will need to be discriminated by future generation experiments able to measure signatures of primordial non-Gaussianities~\cite{Gangui:1993tt,Acquaviva:2002ud,Maldacena:2002vr,Babich:2004gb,bel}, which are sensitive to the specific properties of the considered inflationary model (see, e.g. ~\cite{Bartolo:2004if,Chen:2010xka,Arkani-Hamed:2015bza}).
It is well known that the quest for non-Gaussian signatures requires the study of equations beyond the canonical first-order perturbative approximation, based on considering small metric fluctuations around the Friedmann-Lema\^itre-Robertson-Walker (hereafter FLRW) background. In principle, this perturbative approach is the most suitable one in order to give high precision phenomenological predictions, however, dealing with at least second-order perturbative equations can be really tricky. This difficulty led a non-negligible part of the scientific community 
to search for alternative methods which are able to overcome the difficulty of manipulating higher-order equations ~\cite{Abolhasani:2018gyz,Abolhasani:2013zya,Wands:2010af}. In this note we will focus on a reanalysis of the so-called $\Delta N$ formalism ~\cite{Sasaki:1995aw,Lyth:2004gb,Wands:2000dp,Sasaki:1998ug,Dias:2012qy,Sugiyama:2012tj}, pointing out some  issues in its implementation.
Usually, in the $\Delta N$ formalism the metric is taken in the ADM
decomposition performing a  gradient expansion instead of the conventional perturbative one based on small deviations from a homogenous
background metric. The key quantity is the local number of e-folds ${\cal N}$, that can be defined as the integral of the expansion of a velocity
field defined in terms of an initial and final hyper-surface ${\cal S}$. A
certain level of ambiguity exists on the choice of ${\cal S}$~\cite{Lyth:2004gb,Sugiyama:2012tj,Suyama:2012wi,Garriga:2015tea}
and the role of the initial hyper-surface ${\cal S}_0$ has been often
overlooked. 
An alternative approach to the $\Delta N$ formalism was based on the current $\zeta_\mu$, proposed some years ago 
by Langlois and Vernizzi~\cite{Langlois:2005qp,Langlois:2006vv,Langlois:2006iq,Langlois:2008vk,Langlois:2010vx}
as a suitable non-perturbative generalization of the scalar quantity
$\zeta$~\footnote{Notice that such an approach is different from
  the formalism reviewed in length in the textbook~\cite{ellis_maartens_maccallum_2012}}. 
The link between these two approaches is the local number of e-folds which enters
in both the definition of the $\Delta N$ formula and the current
$\zeta_\mu$. The equivalence of $\zeta_\mu$ and $\Delta N$
formalisms was claimed in~\cite{Suyama:2012wi,Naruko:2012um}. Although a close relation between  $\zeta_\mu$ and $\Delta N$, because of the
role of ${\cal N}$, is not surprising, a full equivalence of these two approaches is far from
being trivial. Indeed, while $\zeta$ of the standard $\Delta N$ formula is
only conserved at super-horizon scales,  it has a nontrivial sub-horizon dynamics. On the contrary,  $\zeta_\mu$ is 
conserved at all scales, in the sense that its Lie derivative along
the flow is exactly zero in the adiabatic case.
 Therefore an equivalence between the two approaches is hardly
 achievable. 

The outline of the paper is as follows. In Section
\ref{e-folding_sec} the local number of e-folds
${\cal N}$ is defined and computed up to first order in perturbation
theory. Section~\ref{nonpert} is devoted to the study of the
currents $\zeta_\mu$ and ${\cal R}_\mu$ introduced in~\cite{Langlois:2005qp},
reconsidering their relation with the standard curvature perturbation of constant density
hyper-surfaces $\zeta$ and the comoving curvature perturbation ${\cal
  R}$. In Section \ref{DeltaN_sec}, the influence of the choice of the class of
hyper-surfaces on the $\Delta\, N$ formula and the relation with the current $\zeta_\mu$ is
reanalysed.  Finally, in
Section~\ref{scalar} we consider in detail as an example a scalar field. Our main conclusions are drawn in Section ~\ref{Conc_sec}.

\section{The local number of e-folds }
\label{e-folding_sec}

\begin{figure}[ht!]
\begin{center}
\includegraphics[width=8cm]{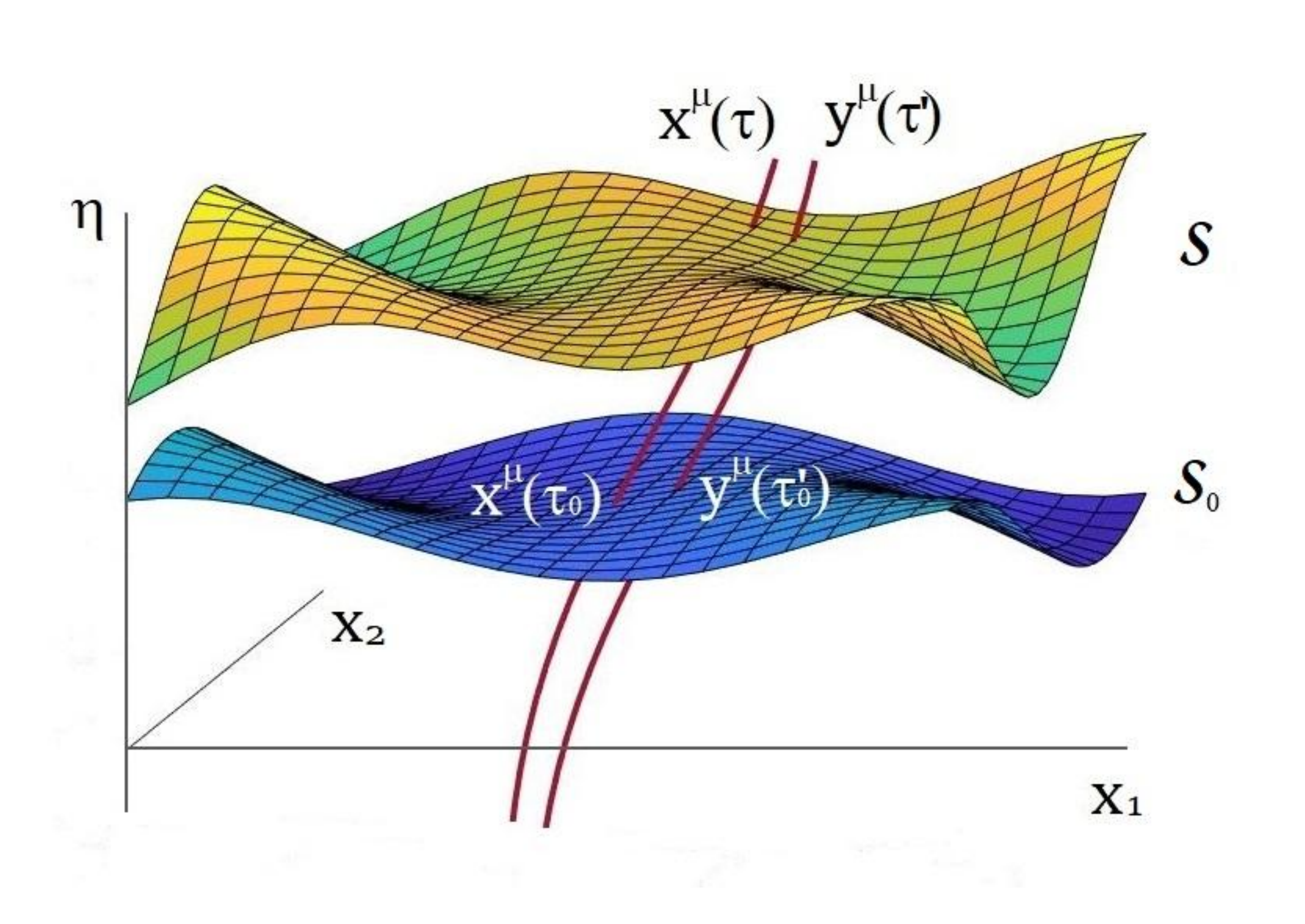}
\caption{The congruence of $u^\mu$ which intersects two
  generic space-like initial and final hyper-surfaces. Once the vector
  field is fixed, the point  $x^\mu$ on the final hyper-surface is uniquely
  determined by the initial point  $x_0^\mu$ on the initial hyper-surface.}
\label{sur}
\end{center}
\end{figure}

One of the main physical quantities crucial for the $\Delta N$ formula, is the so-called local number of e-folds ${\cal N}\left(\eta,x^i\right)$, which
generalizes the number of e-folds in a de Sitter spacetime. We consider a perturbed FLRW universe with metric $g_{\mu \nu}$ in the presence of a
perfect fluid, with 4-velocity $u^\mu$; focusing on scalar modes only, at linear order in perturbation theory we have
\begin{equation}
\begin{split}
& g_{00}= - a^2 (1+2 \,  A) \, , \qquad  g_{0i}=a^2 \, \partial_i B \, , \qquad
g_{ij}=a^2 \, \gamma_{ij}\left( 1-2 \, \psi \right)+ 2 \,a^2 \, \partial_i \partial_j E \, ,\\
& \qquad \qquad \qquad \qquad \qquad u_{\mu}= a \left(-(1+A), \, \partial_i v \right) \, ;
\end{split}
\label{pertg}
\end{equation}
where latin indices run from $1$ to $3$ and are raised/lowered with the unperturbed 3-metric $\gamma_{ij}$, while greek indices are used
to describe space-time coordinates. Notice that the time coordinate $x^0\equiv \eta$ is the background conformal time and
the perturbed metric is given in a generic gauge. In general, by a suitable choice of gauge, only three out of the five scalars correspond to physical degrees of freedom.
An important physical quantity, often used in this paper is the volume
expansion scalar, defined as the 4-divergence of the 4-velocity: 
\begin{equation}
\theta= \nabla^{\mu} u_\mu= 3 \, \dot{{\cal N}} \, ;
\end{equation}
where, in general, for any scalar quantity $f$ we define  $\dot{f} = u_\mu \nabla^{\mu} f= \frac{d f}{d\tau}$, where $\tau$ is the fluid's
proper time. The local number of e-folds ${\cal N}$ can be defined by integrating the expansion
$\theta$ of the fluid velocity along its world-lines; namely
\be
{\cal N}(\eta, x^i) = \frac{1}{3}\int_{{\cal S}_0}^{\cal S} \theta  \, d \tau = -\frac{1}{3} \int_{{\cal S}_0}^{\cal S}  \frac{d\rho}{(\rho+p)} \, ;
 \ee
the congruence of $u^\mu$ is supposed to pierce the hyper-surfaces ${\cal S}_0$ and ${\cal S}$ only once.  Finally, in the last relation we have used the energy-momentum
tensor (EMT) conservation of the fluid
\begin{equation}
\label{Con_par}
u^{\nu}\nabla^{\mu} T_{\mu\nu}=-\dot{\rho}-\theta \,
\left(\rho+p\right)=0 \, .
\end{equation}
Being the final point of the
world-line chosen to be $x^\mu=(\eta,x^i)$ and the congruence fixed,
the point $x_0^\mu \, \in \, {\cal S}_0$ is uniquely determined by tracing
back the world-line until it intersects ${\cal S}_0$, as shown in
figure \ref{sur}. In general, changing $x^i$ is equivalent to changing $x^i_0$ and the world-line which crosses ${\cal S}_0$.

At the linear order, see Appendix~\ref{first-zeta-sec}, one gets
\be
\begin{split}
{\cal N}&= \int_{wl} d\eta' \, a\left[1+A +\cdots
\right]\left[\bar{\theta}+\theta^{(1)} + \cdots\right]\\
&=\ln\left(\frac{a(\eta)}{a(\eta_0)}\right)+ \frac{1}{3} \nabla^2
\left(E|_{\left(\eta_0,x^i\right)}^{(\eta,x^i)}+\int_{\eta_0}^\eta \,d\eta' \,
 \left(v-B\right) \right)-
\psi|_{\left(\eta_0,x^i\right)}^{(\eta,x^i)} \;
\end{split} 
\ee 
where $\bar{\theta} = 3 a'/a$ is the background value of $\theta$. We
stress that there is a one-to-one relationship between the final
point $x^\mu=(\eta,x^i)$ in {\cal S} and $x_0^\mu \, \in \, {\cal S}_0$. 
It is natural to impose that the initial and final space-like hyper-surfaces
are homogenous at the background level.
If the hyper-surface is defined as $f=const.$, where $f$ is a 4-dimensional
scalar function, the following relation holds
\be 
\partial_\beta (f|_{{\cal S}_0})= \partial_{\mu_0} (f |_{{\cal S}_0}) \, \partial_\beta x_0^\mu \, ,
\ee
and according to our
assumptions $f= \bar f(\eta) + f^{(1)}(\eta,x^i)$ and we have in
perturbation theory 
\be
\partial_i (\eta |_{{\cal S}_0})=-\left(\frac{\de_i  f^{(1)}}{\bar f'}\right)|_{ 0}+O(2) \, , \qquad \partial_\eta (\eta |_{{\cal S}_0})= O(2) \, .
\label{par}
\ee
where with $0$ we denote a generic point of ${\cal S}_0$. 
Therefore
\be
\label{eta}
\eta_0=\bar \eta_0 -\frac{f^{(1)}}{\bar{f}'}|_{ 0} \, .
\ee
Notice that in the seemingly background term
$\ln\left(\frac{a(\eta)}{a(\bar \eta_0)}\right)$
a non-trivial spatial dependence due to $\eta_0$, which can be further
expanded to give
\be 
\label{fin_e_folds}
\begin{split}
{\cal N}&=\ln\left(\frac{a(\eta)}{a(\bar{\eta}_0)}\right)+{\cal
  H}(\bar{\eta}_0)  \frac{f^{(1)}}{\bar{f}'}|_{ 0} + \frac{1}{3} \nabla^2
\left(E( \eta, x^i)+\int_{\eta_0}^\eta \,d\eta' \, \left(v-B\right)\right)\\
& \qquad \qquad \qquad \; \;-\psi(\eta, x^i)+\left(\psi-\frac{1}{3}\nabla^2 E\right)|_{0}  \, .
\end{split}
\ee
is present. By construction ${\cal N}$ trasforms as the perturbation of
a spacetime scalar function with a time dependent background; namely
\be
{\cal N}^{(1)} \to {\cal N}^{(1)} - \delta x^0 \, \de_\eta \bar {\cal N} \, ,
\qquad \qquad \bar {\cal N} =\ln\left(\frac{a(\eta)}{a(\bar{\eta}_0)}\right) \, .
\ee
Moreover, the two hyper-surfaces needed in the definition of ${\cal
  N}$ are defined in terms of a spacetime scalar $f=$const.
For instance, in the case ${\cal S}_0$ corresponds to a constant-proper-time of the fluid,
the initial hyper-surface will be of the form $\eta_0= \bar
\eta_0+v(\bar \eta_0,x)$, see Appendix \ref{x0}, where $\bar \eta_0$
is the arbitrary value of the conformal time in the gauge where the
spatial velocity of the fluid $v$ is set to zero. 
\section{ Non-perturbative Currents}
\label{nonpert}
In this section we will re-examine the non-perturbative currents $\zeta_\mu$ and ${\cal R}_\mu$ 
introduced in~\cite{Langlois:2005qp} by Langlois and Vernizzi and the relation with the constant-$\rho$ curvature perturbation~\cite{Bardeen:1983qw,Wands:2000dp} (for a review see for
instance~\cite{Malik:2008im})
\be
\zeta = -\psi +
\frac{\rho^{(1)}}{3(\bar \rho + \bar p) } \, ,
\ee
and the gauge-invariant comoving curvature perturbation~\cite{Bardeen:1980kt,Lyth:1984gv} 
\be 
{\cal R} = -\psi + {\cal H} \, v  \,.
\ee
See also ~\cite{Rigopoulos:2003ak} for an
alternative definition of vectors related to $\zeta$ or ${\cal R}$ at the non-linear level and 
their compatibility with the $\Delta N$ formalism, after a suitable initial hyper-surface choice~\cite{Rigopoulos:2005us,Tzavara:2010ge}.   

\no
Given a fluid with four-velocity $u^\mu$, $\zeta_\mu$ is defined as
\be
\zeta_\mu  
=\partial_\mu {\cal N}-\frac{\dot{{\cal N}}}{\dot{\rho}}
\partial_\mu \rho \, ,
\label{zetadef}
\ee
where ${\cal N} $ is precisely the local number of e-folds
computed between two generic hyper-surfaces, as
discussed in the previous section.  The quantity $\zeta_\mu$ is
fully non-perturbatively defined and, in the case of an adiabatic fluid, one can show
that ~\cite{Langlois:2005qp} it does not change along the fluid lines, in
other words, its Lie derivative along $u^\mu$ vanishes 
\be
\label{lie_cons}
{\cal L}_u \, \zeta_\mu=  0 \, . 
\ee 
The previous relation can be considered as a non-perturbative
conservation law for $\zeta_\m$ and is valid at any scale. 
In the case of a barotropic and irrotational perfect fluid $\zeta_\mu$ can be computed exactly, showing that it
depends only on the choice of ${\cal S}_0$.
By using the $\theta$ definition and Eq. (\ref{Con_par}) one gets
\begin{equation}
\zeta_\mu= -\frac{1}{3} \partial_\mu \int_{\tau_0}^{\tau} d\tau'
\frac{\dot{\rho}}{(\rho+p)}+\frac{1}{3}\frac{\partial_\mu\rho}{(\rho+p)}
\, ;
\end{equation}
or alternatively, by introducing the 1-form ${\pmb \chi}$
\be
\pmb{\chi}= \chi_\mu \, \pmb{dx^\mu} = \frac{\partial_\mu \rho}{( \rho
  +p)} \, dx^\mu \, , 
\ee
as a 1-form $\pmb\zeta$ whose components are given by
\be
\label{zeta_mu}
\zeta_\mu = \frac{1}{3} \left( \chi_\mu - \partial_\mu \int  \chi_\nu dx^\nu 
\right) \, .
\ee
For a barotropic fluid, for which  $p=p(\rho)$, the 1-form $\pmb{\chi}$ is closed, namely
\be
\label{baro}
 \pmb{d \chi} =- \frac{\pmb{d } \rho \wedge \pmb{d} p}{( \rho +p)^2} =0
\, ;
\ee
equivalently, in components, $\de_{[\mu} \chi_{\nu]} =0$. By using the Poincar\'e lemma, one can find, at least locally, a function $\beta$
such that $\chi_\mu=  \partial_\mu \beta$.
One can get
\be
\beta(\rho) = \int^\rho \frac{dx}{x+p(x)} \, , \qquad \pmb{d} \beta =
\pmb{\chi} \, .
\label{beta1}
\ee
Thus, neglecting any topological complication,  for a barotropic fluid
we can compute $\zeta_\mu$, exactly arriving at the following simple expression
\be
\zeta_\mu =\frac{1}{3} 
\left[ \partial_\mu \beta - \partial_\mu \left(\beta -
    \beta_0 \right) \right] = \frac{\partial_\mu \beta_0}{3} =\frac{\partial_\mu \rho |_{0}}{3(\rho+p) |_{0}} \,   .
\label{barres}
\ee
where $0$ indicates that the relevant quantity is evaluated on the initial 3-surface ${\cal S}_0$. One can easily see that the Lie derivative
along $u$ of Eq. (\ref{barres}) is given  by
\be
{\cal L}_u \pmb\zeta = \pmb{d} \left(u^\mu \de_\mu  \beta_0 \right)=0 \, ,
\label{npres}
\ee
being, by definition, $\beta_0$ evaluated on ${\cal S}_0$; the above result is in agreement with (\ref{lie_cons}). As a result, in the
case of a barotropic fluid, (\ref{barres}) shows that $\zeta_\mu$ depends exclusively on the initial hyper-surface ${\cal S}_0$ and in
this sense it is trivial as a dynamical quantity. An alternative
interpretation of (\ref{barres}) is that for a barotropic fluid,  when the local number of e-folds is computed on a
constant $\rho$ hyper-surface, then $\zeta_i \equiv 0$; such a result is non-perturbative.

A non-trivial dynamics is reintroduced when the fluid is non-barotropic, namely when
\be
\Gamma_\mu = \de_\mu p - c_s^2 \, \de_\mu \rho \neq 0  \, , \qquad c_s^2 =
\frac{\dot{p}}{\dot{\rho}}  \, .
\label{Gamma}
\ee
In this case $\zeta_\mu$ not only depends on the final hyper-surface but also on the world-line.

This exact result (\ref{barres}) can be used as an alternative starting point for a perturbative expansion. We note that, for any function $f$,
say $\rho$, defined in ${\cal S}_0$ where $\tau= \tau_0$ we have
\be
\begin{split}
\partial_\mu \rho(\eta_0,x_0^i)&= \frac{\partial\rho(\textbf{x}_0)}{\partial x_0^\nu}\frac{\partial x_0^\nu }{\partial x^{\mu}}\\
&=\frac{d \bar{\rho}(\eta_0)}{d \eta_0} \frac{\partial
  \eta_0}{\partial x^{\mu}}+\frac{\partial \rho^{(1)}|_0}{\partial
  x_0^j}\frac{\partial x_0^j}{\partial x^{\mu}}\;  ;
\end{split} 
\ee
in ${\cal S}_0$ the conformal time will be a function of $x^i$. Expanding, we have $x^i=x_0^i+O(1)$ and $\frac{\partial
  x_0^i}{\partial \eta}\approx \frac{\partial x^i}{\partial
  \eta}=O(1)$. Furthermore, using the relations
$\eta_0^{(1)}=-\frac{f^{(1)}}{\bar f}\left(\bar{\eta}_0,x^i\right)$ and $\partial_\eta
\eta_0= O(2)$ (see Section \ref{e-folding_sec}), we find  for the first-order
expansion $\zeta^{(1)}{}_\mu$ of $\zeta_\mu$
\be
\begin{split}
\label{gen_f}
&\zeta_0^{(1)} =0 \, , \qquad \qquad \zeta_i^{(1)} =\partial_i
\zeta_{s}^{(1)}  \, ,\\[.2cm]
&\zeta_{s}^{(1)}= \frac{{\cal H}(\eta_0)}{\bar{f}'(\eta_0)} \, f^{(1)}\left(\bar{\eta}_0,x^i\right)-\frac{{\cal H}(\eta_0)}{\bar{\rho}'(\eta_0)}\rho^{(1)}(\bar{\eta}_0,x^i) \, .
\end{split} 
\ee
From (\ref{gen_f}) and (\ref{npres}) it is clear that, for a barotropic fluid, $\zeta_s^{(1)}$ is
defined on the initial hyper-surface; any dependence on the final 
hyper-surface cancels out and thus it is not related to $\zeta$~\cite{Langlois:2005qp} and to the $\Delta N$
formalism~\cite{Naruko:2012um,Suyama:2012wi}. Such argument is valid
at any scale.  Notice that $\zeta_s^{(1)}$ is gauge invariant.

Finally, as an additional check, $\zeta_\mu$ can be computed perturbatively starting from its definition
(\ref{zetadef}). By using the results of Appendix \ref{first-zeta-sec}, at the linear order
and for a generic perfect fluid, we have that 
\bea
&&\zeta_0^{(1)} =0 \, , \qquad \qquad \zeta_i^{(1)} =\partial_i
\hat{\zeta}_{s}^{(1)}  \, ,\\[.2cm]
&&\hat{\zeta}_{s}^{(1)}=\zeta+\frac{1}{3}\left(\nabla^2 E+\int_{wl}
  d\eta \, \nabla^2 (v-B) \right) +\left[ \psi +
{\cal H} \,
\frac{f^{(1)}}{\bar{f}'}-\frac{1}{3}\nabla^2
E \right]_{|(\bar\eta_0,x^i)}\, .
\label{zetaLVp}
\ea
Where we suppose to find a scalar function $\hat{\zeta}_s^{(1)}$ such
that $ \zeta_i^{(1)} \equiv \de_i\hat{\zeta}_s^{(1)}$, also in the
non-adiabatic case.
From the standard relation 
\be
\zeta'=-\frac{1}{3}\nabla^2 \left(E'+v-B\right)- \frac{{\cal H}}{\bar{\rho}+\bar{p}}\Gamma^{(1)} \, ,
\ee
we can write (\ref{zetaLVp}) as
\be
\begin{split}
\hat \zeta_s^{(1)} &= \zeta - \int_{\eta_0}^{\eta} d \eta' \, \left[ \zeta'  +
\frac{{\cal H}}{\bar{\rho}+\bar{p}}\Gamma^{(1)} \right] +\psi\left(\bar\eta_0,x^i\right) +
{\cal H} \,
\frac{f^{(1)}}{\bar{f}'}\left(\bar\eta_0,x^i\right) \\
&=\zeta\left(\bar\eta_0,x^i\right) +\psi\left(\bar\eta_0,x^i\right) +
{\cal H} \,
\frac{f^{(1)}}{\bar{f}'}\left(\bar\eta_0,x^i\right) - \int_{\eta_0}^{\eta} d \eta' \,  
\frac{{\cal H}}{\bar{\rho}+\bar{p}}\Gamma^{(1)}  \, .
\end{split}
\label{zetaLV}
\ee
Let us stress that $\hat{\zeta}_{s}^{(1)}$ is a gauge invariant
quantity which depends  only on the choice of the initial and final hyper-surface.
If $\Gamma^{(1)}=0$ (barotropic case), $\hat{\zeta}_{s}^{(1)}\equiv \zeta_s ^{(1)}$ is constant in time at {\it all scales} and coincides with
its value at $\eta=\eta_0$, namely only on the choice of the initial hyper-surface, confirming the result (\ref{gen_f}) based on the exact expression (\ref{barres}).
It is worth to point out that, in order to get (\ref{zetaLV}), the somehow hidden dependence on
$x^i$ of the $\eta_0(x)$, parametrizing the initial hyper-surface,
gives the term of the form  ${\cal H} \,
\frac{f^{(1)}}{\bar{f}'}$. The lesson is that $\zeta_\mu$ as 
dynamical quantity  is  trivial in the barotropic case. Moreover, if the
hyper-surface $f$ is taken to be  a uniform density hyper-surface,
namely $f=\rho$, then $\zeta^{(1)}_\mu=0$. Summarizing
\begin{itemize}
\item $\zeta_i^{(1)} \neq \de_i \zeta$;
\item For an adiabatic fluid, $\zeta_i^{(1)}$ is conserved  at all scales at the first order in perturbation theory and does not depend on the choice of the final hyper-surface.
\item
  $\zeta_\mu^{(1)}=0$ when a uniform density hyper-surface is considered.
\item
  $\zeta_i^{(1)}=\de_i\zeta(\eta_0,x^i)$ when the initial hyper-surface is that
  $f =\eta$ and $\psi=0$, namely a flat slice is considered.
\end{itemize}
In Appendix \ref{sec_zeta} we have computed
$\zeta_\mu$ at second order in perturbation theory, starting
from Eq. (\ref{barres}), verifying that (\ref{lie_cons}) holds. \\
Besides $\zeta_\mu$, it is possible to define another quantity, ${\cal
  R}_\mu$, related to the curvature of comoving hyper-surfaces,  
defined as~\cite{Langlois:2005qp}
\be
{\cal R}_\mu=h_\mu^\nu  \de_\nu {\cal N}= \zeta_\mu - \frac{D_\mu \rho}{3(p+\rho)} \, , \qquad
D_\mu \rho = h_\mu^\nu \de_\nu \rho \, ;
\ee
where $ h_{\mu \nu}= g_{\mu \nu} +u_\mu \, u_\nu$ is the projector orthogonal to $u_\mu$. 
In the barotropic fluid case, while $\zeta_\mu$ is exactly
conserved at all scales, this is not the case for ${\cal R}_\mu$.
Interestingly, the difference between ${\cal R}_\mu$
and $\zeta_\mu$ can be written as
\be
{\cal R}_\mu - \zeta_\mu=
\frac{\theta}{3\dot{\rho}}\left(\partial_\mu \rho+\dot{\rho}\, u_\mu\right)
\equiv  \frac{\theta}{3 \, \dot{\rho}} D_\mu \rho\, ,
\label{Rdef}
 \ee
where $D_\mu \rho$ can be red as a covariant and non-linear generalization of the comoving density perturbation. 
As before, it is also possible to find a
perturbative expansion for ${\cal R}_\mu$. The first non-trivial order
is the linear one, for which by perturbing (\ref{Rdef}) and by using
(\ref{gen_f}), one gets 
\bea
&& {\cal R}_0 
= \bar{\cal N}'+\frac{1}{3} \bar\theta \, \bar u_0+ {\cal
  N}'{}^{(1)}+\frac{\bar u_0 \, \theta^{(1)}}{3}+\frac{u_0^{(1)}}{3}\bar \theta 
= O(2) \, \\[.2cm]
&& {\cal R}_i=\de_i {\cal R}_s\, , \qquad {\cal R}_s= \zeta_{s}-\int_{\eta_0}^{\eta}
\frac{{\cal H}}{\bar{\rho}+\bar{p}}\Gamma^{(1)} \, d\eta\,+\frac{{\cal H}}{\bar\rho'}\Delta\rho=\hat{\zeta}_{s} +\frac{{\cal H}}{\bar\rho'}\Delta\rho\, ,
\ena 
where $\Delta\rho=\bar{\rho}'  \, v+\rho^{(1)}$ is the so-called comoving-gauge density perturbation. 

Moreover,
by the above analysis it is clear that when the ${\cal R}$ modes are not conserved, being $\Delta \rho \propto {\cal R}-\zeta$, the difference ${\cal
  R}_\mu-\zeta_\mu$ can be used as a tool for the study of the
Weinberg theorem~\cite{Weinberg:2003sw} at the non-perturbative level even for a fluid; see, for a
recent
discussion~\cite{Kinney:2005vj,Motohashi:2014ppa,Chen:2013kta,Akhshik:2015nfa,Celoria:2017xos}. We leave the study of such violations for future work.\\
Hereafter, let us give a concrete and natural example for the choice
of  ${\cal S}_0$, taking a constant-proper-time
hyper-surface. In this case we can replace $-\frac{f^{(1)}}{\bar f}$ with the scalar velocity $v$ (see Appendix \ref{x0})
\be 
\label{gauge}
\begin{split}
\zeta_{s}^{(1)}&=-\frac{{\cal H}(\eta_0)}{\bar{\rho}'(\eta_0)}\left[\bar{\rho}'(\eta_0)v(\bar{\eta}_0,x^i)+\rho^{(1)}(\bar{\eta}_0,x^i)\right] \\
&\equiv-\frac{{\cal H}(\eta_0)}{\bar{\rho}'(\eta_0)}\Delta\rho(\bar{\eta}_0,x^i)\, ,
\end{split}
\ee
Note that,  in the Fourier space, when a constant-proper-time initial
hyper-surface is used, for super-horizon modes, in the adiabatic case
\be 
{\cal R}_s \,, \zeta_s \,\longrightarrow \, 0 \, \qquad
\text{super-horizon} \, ; 
\ee 
where we have used that from the perturbed Einstein equations $\Delta\rho \sim k^2 \psi$. The
results are  different from~\cite{Langlois:2005qp}.

In conclusion, the currents $\zeta_\mu$
and ${\cal R}_\mu$ are not in general directly related to $\zeta$ or ${\cal R}$, which are conserved in the super-horizon limit for an adiabatic fluid.

\section{\texorpdfstring{$\Delta N$}{Lg} Formula and Relation with  \texorpdfstring{$\zeta_\mu$}{Lg}}
\label{DeltaN_sec}
As we have seen in the previous section the integral of the expansion
$\theta$, at small momenta (large distances) involves the gravitational
potential $\psi$ of the perturbed metric (\ref{pertg}); the $\Delta \, N$ formula exploits this integral by computing 
how the local number of e-folds changes moving along two different hyper-surfaces. According
to the {\it Separate Universe} approach ~\cite{Lyth:2004gb,Wands:2000dp},
perturbation theory can be formulated as a derivative expansion and
terms with more than one spatial derivative can be neglected at large distances.
Following~\cite{Starobinsky:1982ee,Salopek:1990jq,Sasaki:1995aw,Lyth:2004gb,Lyth:2005fi},
by choosing a suitable initial and final hyper-surfaces
${\cal S}_0$ and ${\cal S}$, one can isolate the perturbation mode $\psi$, which is
proportional to the curvature of the $\rho$=constant hyper-surface and
is conserved on super-horizon scales. A number of recipes for the
choice of ${\cal S}$ have been proposed in the literature. For
instance, in~\cite{Lyth:2004gb}, one first computes ${\cal N}_A$
taking  ${\cal S}_0$ to be a flat constant-conformal-time hyper-surface, while ${\cal S}$ is a  slicing with $\rho^{(1)}$=0 hyper-surface, then the same quantity, 
${\cal N}_B$, is computed, taking both ${\cal S}_0$ and ${\cal S}$
as flat constant-conformal-time hyper-surfaces; finally the $\Delta \,
N$ formula is defined as ${\cal N}_A-{\cal N}_B$; notice that the
additional hypothesis that the two initial hyper-surfaces are tangent
in the point of interest $x^i$. From our general expression (\ref{fin_e_folds}), the role of ${\cal N}_B$ is that of subtracting the
background value $\bar {\cal N}$ to single out $\psi$. By
Eq. (\ref{fin_e_folds}), we get
\be
\begin{matrix}
{\cal N}_A = \bar {\cal N}  -\psi(\eta,x^i) \\[.4cm]
{\cal N}_B = \bar {\cal N} \end{matrix} \qquad 
\Rightarrow \qquad \Delta N=  -\psi(\eta,x^i) \, .
\label{sas}
\ee
Alternatively, according to~\cite{Sugiyama:2012tj}, one should define $\Delta N$ as the 
  difference of ${\cal N}_A$, computed by using a flat conformal
  hyper-surfaces and ${\cal N}_B$  computed by gauge transforming
  ${\cal N}_A$ from the flat to the uniform density gauge, but only on the final hyper-surface. The result is the same of Eq.(\ref{sas}).
On the other hand,  ${\cal N}$, obtained by taking two
constant-conformal-time hyper-surfaces and neglecting spatial
derivatives (Separate Universe assumption),  is given by (see Eq. (\ref{fin_e_folds}))
\be
{\cal N}=\bar {\cal N}-\psi(\eta,x^i)+\psi(\bar\eta_0,x^i)
\, \, ,
\label{amb}
\ee
and typically  the contribution of  $\psi(\eta_0,x^i)$ is neglected.
Notice however that such a term  is important; indeed, in the gauge
$\rho^{(1)}=0$, the energy-momentum conservation leads to $\psi'=0$
and then $\Delta N=0$. 
Other definitions can be found,  see for instance~\cite{Suyama:2012wi,Garriga:2015tea}. 

As a final comment we point out that, 
although Eq.(\ref{fin_e_folds}) can be extended to
hyper-surfaces $\eta=$const., by setting $f^{(1)}=0$, such a
choice is ambiguous, being the surface defined by using the unperturbed
coordinated time $\eta$ of a perturbed universe and $\psi$ is in an 
unspecified gauge. The choice of $\eta=$const. in a perturbed universe does not
identify uniquely the metric perturbations; indeed, by an infinitesimal change
of coordinates the metric takes a physically equivalent form, leaving two
scalars to be gauge fixed.

Let us now compare $\zeta_\mu$ with  $\Delta N$.
In~\cite{Suyama:2012wi,Naruko:2012um} the former  was claimed to be
equivalent to the $\Delta N$ formalism. The starting point is the
relation $\zeta^{(1)}_\mu=(0,\partial_i \zeta)$, which we have
shown that is not correct. According to their reasoning, by
using the same hyper-surfaces for the computation of $\Delta N$, they find
\be
\zeta_i= \partial_i \left(\psi \left(\eta, x\right)-\psi\left(\bar \eta_0,x\right)\right) \equiv \partial_i \Delta N \, , 
\ee
However, this result is based on the results presented in \cite{Langlois:2005qp} obtained without taking
into account the subtle issues previously analyzed, and it is not
coherent with  Eq. (\ref{gen_f}) or with the general extension for non-barotropic fluids,  
Eq. (\ref{fin_zeta}) (see Appendix
\ref{first-zeta-sec}).

Indeed, from the above analysis it is clear that $\zeta_\mu$, and the related quantity 
$\zeta_{s}^{(1)}$ are conceptually
rather different from the constant-$\rho$ curvature perturbation
$\zeta$. Indeed, in the barotropic case, $\zeta_\mu$ is a quantity
that depends on the initial hyper-surface only as we have shown in
Eq.(\ref{barres}), while on the contrary, by construction $\Delta N$
is sensitive to the final hyper-surface, see Eq.(\ref{sas}). At the
linear order in perturbation theory this shows up from the fact that 
$\zeta_{s}^{(1)}$ is a function of the spatial coordinates only, {\it before} the super-horizon limit 
is taken and thus no 
genuine sub-horizon dynamics is present, in sharp contrast with $\zeta$. A rather formal
comparison can be made by  choosing the initial hyper-surface for the
computation of $\zeta_\mu$ to be the same as the one used for the
$\Delta N$ computation; namely we set in (\ref{gen_f}) $f^{(1)}=0$ and
  take $\psi(\eta_0,x^i)=0$, thus we get
  \be
  \zeta_s^{(1)} = \zeta(\eta_0,x^i)   \qquad \qquad \eta=\eta_0 \text{
    and flat} \, .
  \ee
On the other hand, for the final uniform density hyper-surface
in~(\ref{sas}) we have that $\Delta N= -
\psi(\eta,x^i) \equiv \zeta(\eta,x^i)$. Thus, if we are
interested in super-horizon scales only, being $\zeta$ conserved, we get the
somehow accidental relation
\be
\zeta_s^{(1)} = \Delta N \, .
\label{equiv}
\ee
In spite of the previous relation,  the two objects are intrinsically
different. Taking a standard scenario with adiabatic
initial conditions, $\zeta_\mu$ is completely determined at all scales
by its initial value when  inflation starts. On the contrary, $\Delta
N$ has a non-trivial dynamics and is constant only at the
zero order of the gradient expansion $(k \longrightarrow 0)$.\\
We point out that the relation $\zeta_i =\partial_i \zeta_{s}^{(1)}$
can be extended beyond perturbation theory thanks to the non-perturbative nature of
(\ref{barres}); indeed for a barotropic fluid, choosing in addition a
set of adapted coordinates such that $u^\mu=(u^0, 0)$ (comoving
threading) we have that
\be
\zeta_i= \frac{1}{3}\frac{\partial_i \rho}{(\rho+p)}|_{(\eta_0, x^i)}=\frac{1}{3} \partial_i \left(\int_{\rho(\eta_0,x^i_{in})}^{\rho(\eta_0,x^i)}\frac{dx}{x+p(x)}\right) \, , 
\label{scalnp}
\ee 
which is {\it completely non linear} and time independent {\it at all scales}, in sharp contrast with the
$\Delta N$ formalism. In general, (\ref{scalnp})  does not hold for a
generic choice of threadings where  $x_0^i \neq x^i$.  

\section{Scalar Field}
\label{scalar}
As an explicit example let us consider a scalar field. One should
first emphasize that a scalar field and a
perfect fluid are inequivalent when thermodynamics is taken into
account, unless a shift symmetry is
present~\cite{Ballesteros:2016kdx}. For a real scalar field with a
canonical kinetic term and a potential $V(\phi)$, we have
\be
\begin{split}
& \rho=K+ V \, , \qquad p=K -V \, , \\
&K = -\frac{1}{2} g^{\mu\nu} \partial_\mu \phi \partial_\nu \phi \, ;
\end{split}
\ee
the velocity is
\be
u_\mu= - \frac{\partial_\mu \phi}{\sqrt{-g^{\alpha\beta}\partial_\alpha \phi \partial_\beta \phi}} \, .
\ee
Of course, in general the relation between $p$ and $\rho$ will be
non-barotropic, indeed $p=\rho-2 \, V$; as a result, the quantity
$\Gamma_\mu$, see (\ref{Gamma}),  which measures that effect will be non-zero
and given by~\cite{Langlois:2006vv}
\be
\Gamma_\mu 
=\left(1-\frac{\dot{p}}{\dot{\rho}}\right)D_\mu \rho-2D_\mu \, V 
=2 \, V_\phi \, \frac{\dot{\phi}}{\dot{\rho}}D_\mu \rho \, ;
\ee
where we have used that $D_\mu \phi=D_\mu V=0$;  in particular, setting $\partial
K/\partial (\de_\mu \phi) = K^\mu$, we arrive at the following
expression
\be 
\dot{\rho}=K^\mu  \, {\cal L}_u \partial_\mu \phi+V_\phi \, \dot{\phi} \, ,
\ee
Expanding at linear order around a homogeneous cosmological
background for which $\phi = \bar \phi(t) + \phi^{(1)} + \cdots$, we
obtain for the linear perturbation of $\Gamma_\mu$
\be
\begin{split}
&\Gamma_0^{(1)}= 0 \, ; \\
& \Gamma_i^{(1)}=\partial_i \Gamma^{(1)} \, , 
\qquad \Gamma^{(1)}=2 \, \bar{V}_\phi
\frac{\bar\phi'}{\bar\rho'}\Delta\rho^{(1)} = - \frac{2 \, \bar{V}_\phi}{3
  \, {\cal H}}\left[ -A\, \bar{\phi }'+ \phi^{(1)}{}'+\phi^{(1)}  \left(3 {\cal H} +\bar{V}_\phi\frac{a^2}{\bar\phi'}\right)
 \right]\, .
\end{split}
\ee
In particular
\be
\Delta \rho =-A\frac{ \bar{\phi }'^2}{a^2}+ \phi^{(1)}{}'\frac{ \bar{\phi }'}{a^2}+\phi^{(1)}  \left(\frac{3 {\cal H} \bar{\phi}' }{a^2 }+\bar{V}_\phi\right)
\ee
For what concerns ${\cal R}_s$, it is given by
\be
{\cal R}_s =\zeta_s-2\int_{\eta_0}^\eta d\eta \, \frac{{\cal H} \, \bar\phi' \,
  \bar{V}_\phi}{(\bar\rho+\bar p)\bar\rho'} \, \Delta\rho +\frac{{\cal
    H}}{\bar \rho'} \, \Delta\rho \, .
\ee
Finally, remember that $\zeta_s$ is given by
\be
\zeta_s =\frac{{\cal H}}{\bar f'}f^{(1)}-\frac{{\cal H}}{\bar \rho'}\rho^{(1)} |_{(\eta_0,x^i)} \, ,
\ee
with $\rho^{(1)}= \Delta \rho-\bar \rho' v^{(1)}=\Delta \rho+\bar \rho'\frac{\phi^{(1)}}{\bar \phi'} $.
Our result differs from the ones found in~\cite{Langlois:2006vv}. Notice that on super-horizon scales 
\be
\zeta={\cal R}= -\psi - \frac{\cal H}{\bar\phi'} \, \phi^{(1)} \, ,
\ee
therefore as shown by this simple example, is quite evident that there is no direct correlation between the currents and the standard curvature perturbations $\zeta$ and ${\cal R}$.
\section{Conclusions}
\label{Conc_sec}
In~\cite{Langlois:2005qp} a generalization of the curvature perturbation of the constant $\rho$ hyper-surfaces $\zeta$ was proposed,
based on a non-perturbative approach. We have computed
$\zeta_\mu$ at the full non-perturbative level in the case of a barotropic fluid, showing that the relation with the standard quantity $\zeta$
is non-trivial. By matching the expansion of our non-perturbative expression for $\zeta_\mu$, we have found that, although
at the linear level, $\zeta_i^{(1)} = \de_i \zeta_{s}^{(1)}$, $\zeta_{s}^{(1)} \neq \zeta$. In particular, while $\zeta$ is time-independent only on
super-horizon scales, the time derivative of  $\zeta_{s}^{(1)} $ vanishes identically. That $\zeta_{s}^{(1)}{}'=0$ at all scales can also be deduced
by expanding ${\cal L}_u \zeta_\mu=0$. These facts profoundly change the physical interpretation of the $\zeta_\mu$ 4-vector, which is conserved
{\it on all scales} and cannot be compared with the gradient of the curvature perturbation $\zeta$, which is instead conserved only on large super-horizon scales.
Similar considerations apply to second order: while  $\zeta_{s}^{(1)}$ is a genuine gauge-invariant quantity likewise $\zeta$, this is not the case for $\zeta_s^{(2)}$. We have also studied the non-perturbative
generalization ${\cal R}_\mu$ of the comoving curvature perturbation
${\cal R}$ proposed in~\cite{Langlois:2005qp}. While at leading
non-trivial order $\zeta_\mu$ has no dynamics, ${\cal R}_i=\de_i
{\cal R}_s$ is a genuine dynamical quantity. However from our analysis
it follows that ${\cal R}_\mu$ is not a suitable non-perturbative
generalization of the comoving curvature perturbation, but the combination ${\cal R}_\mu- \zeta_\mu$ can
be used as a tool for studying the  violation of the Weinberg
theorem.   \\
We have also critically reconsidered the $\Delta N$ formalism and its relation with the covariant vector $\zeta_\mu$ proposed by Langlois and
Vernizzi. Concerning the $\Delta N$ formalism, we
have clarified some ambiguities arising from the choice of the initial and final space-like hyper-surfaces ${\cal S}_0$ and ${\cal S}$, respectively.
Using the prescription defined in \cite{Lyth:2004gb}, we recover the
standard result $\Delta N=-\psi(\eta,x^i)$, which coincides with
$\zeta(\eta,x^i)$ under the condition $\rho^{(1)}=0$. Elaborating on
the result in~\cite{Langlois:2005qp}, where $\zeta_i$ was claimed to
reduce to the spatial gradient of $\zeta$ in the barotropic case, it
was put forward the  equivalence between the current $\zeta_\mu$ and
the $\Delta N$ formalism \cite{Naruko:2012um,Suyama:2012wi} once the
same prescription of spacetime slicing and threading is
applied. According to the our result, bases on both 
perturbation theory and on the novel  exact  expression for
$\zeta_\mu$ in the barotropic case, a number of issues exist. 
\begin{itemize}
\item The current $\zeta_\mu^{(1)}$ depends only on the initial hyper-surface, on the contrary, in the $\Delta N$ formalism the role of such a surface is to avoid any initial contribution,
\item By using a comoving threading and flat initial hyper-surface,
  $\zeta_s^{(1)}$ accidentally reduces to $\zeta(\eta_0,x^i)$, while
  $\zeta_i$ is exactly time independent at all scales.   
\end{itemize}
Hence, we conclude that the $\Delta N$ formalism, which represents a
genuine dynamical quantity  which reduces to $\zeta$ only on
super-horizon scales, cannot be fully equivalent to $\zeta_\mu$.

\acknowledgments 
We thank Misao Sasaki and Gerasimos Rigopoulos for useful
discussions. SM acknowledges partial financial support by ASI Grant
No. 2016-24-H.0.
LP is supported by the research grant "The Dark Universe: A Synergic Multimessenger Approach" number 2017X7X85K under the program PRIN 2017 funded by the Ministero dell’Istruzione, Università e della Ricerca (MIUR).

\appendix
\section{Parametrization of the constant-proper-time hyper-surfaces}
\label{x0}
Given an irrotational fluid with four-velocity $u^\mu$, consider the constant-proper-time $\tau$ hyper-surface ${\cal S}$, with normal vector $u_\mu
\propto \partial_\mu \tau$. 
Suppose that $\tau$ is a differentiable function, that in a given point $\left(\bar\eta, \bar{x}^i\right)$ gives 
\be
\tau \left(\bar\eta, \bar{x}^i\right)-\tau_0=0\, ,
\ee
with $\tau_0$ constant and $\partial_\eta \tau \left(\bar\eta,
  \bar{x}^i\right) \neq 0$. Thus, thanks to the implicit function theorem, we have
\be
\label{Dini}
\frac{\partial \eta_0(x_0)}{\partial x_0^i}=-\frac{\partial_i \tau}{\partial_\eta \tau}|_0=-\frac{u_i}{u_0}|_0 \, ,
\ee
where the subscript $|_0$ denotes the set of points $\left(\eta_0, x_0^i\right)$, which describes the initial surface ${\cal S}_0$. 
Once $u^\mu$ is fixed, there is a one-to-one correspondence between the point 
$x_0^\mu=(\eta_0, x_0^i)$ on the initial constant $\tau$ 3-surface, with the final point taken to be a generic space-time
point $x^\mu$, as illustrated in fig. \ref{sur}. Thus
\be
\eta_0=\eta_0(\eta,x) \, , \qquad x_0^i= x_0^i(\eta,x) \, . 
\ee
and using the rule for the differentiation of composite functions we get
\be 
\partial_i \eta_0= \frac{\partial \eta_0(x_0)}{\partial x_0^j} \,\partial_i x_0^j \, , \qquad \partial_\eta \eta_0= \frac{\partial \eta_0(x_0)}{\partial x_0^j}\,\partial_\eta x_0^j \, .
\ee  
Finally we can substitute Eq. (\ref{Dini}) obtaining
\be
\label{der_eta_0}
\partial_{\mu} \eta_0= -\frac{u_i}{u_0} \partial_{\mu} x_0^i \, . 
\ee 
One can compute $\partial_\mu x_0^{j}$ from the definition of the $u$ congruence
\be
x^{\mu}=x_0^{\mu}+\int_{\tau_0}^{\tau} u^{\mu}\left(\tau'\right)\, d\tau'  \, .
\ee
Therefore~\footnote{$x^j$ and $\eta$ are independent coordinates
therefore $\partial_\eta x^j=0$.}
\be
\begin{split}
& \partial_i x_0^j=\delta_i^j+\partial_i
x_0^{j\,(1)}=\delta_i^j-\partial_i \int_{\tau_0}^\tau u^j \, 
d\tau' \, , \\
& \partial_\eta x_0^j= \partial_\eta x^j-\partial_\eta \int_{\tau_0}^\tau u^j \, d\tau' = -\partial_\eta \int_{\tau_0}^\tau u^j \, d\tau' \,
\end{split}
\ee
Substituting these relations in Eq. (\ref{der_eta_0}) and
expanding up to second order we obtain
\be
\begin{split}
\label{eta_2}
&\partial_i \eta_0^{(1)}=-\frac{u_i^{(1)}}{\bar{u}_0}|_{\bar{0}}=\partial_i \, v|_{\bar 0} \, , \\
&\partial_i \eta_0^{(2)}=-\frac{u_i^{(2)}}{\bar{u}_0}|_{\bar{0}}-\left(\frac{u_i^{(1)}}{\bar{u}_0}\right)'|_{\bar{0}}\eta_0^{(1)}|_{\bar{0}}+\frac{u_i^{(1)}}{\bar{u}_0}\frac{u_0^{(1)}}{\bar{u}_0}|_{\bar{0}}\\
&\;\;\;\;\;\;\;\;\;\;\;+\left[\partial_j\left(\frac{u_i^{(1)}}{\bar{u}_0}\right)|_{\bar{0}} +\frac{u_j^{(1)}}{\bar{u}_0}|_{\bar{0}} \, \partial_i\right]\, \int_{wl} u^{j\,(1)} \,d\tau \, ,\\
&\partial_\eta \eta_0^{(1)}=0 \, , \\
&\partial_\eta \eta_0^{(2)}=-\frac{u_i^{(1)}}{\bar{u}_0}|_{\bar 0} \, \left(\partial_\eta x_0^j\right)^{(1)} \, ,
\end{split} 
\ee
where with the subscript $|_{\bar{0}}$ we label the point $\left(\bar\eta_0,x^i\right)$.

\section{Perturbative computation for \boldmath \texorpdfstring{$\zeta_\mu$}{Lg}}
\subsection{1-st order, generic fluid}
\label{first-zeta-sec}
In this Appendix we will compute perturbatively $\zeta_\mu$ at first order, in the case of a generic perfect fluid, starting directly from
the $\zeta_\mu$ definition (\ref{zetadef}) and verify that it is coherent with the perturbative expansion of (\ref{barres}), when
$\Gamma_\mu=0$ (barotropic fluid). 
By using the definition of the expansion scalar $\theta$, we find at first order 
\begin{equation}
\label{the}
\theta =\nabla_\mu u^\mu    =\bar{\theta}+ \theta^{(1)}+O(2) 
     =\frac{1}{a}\left[ 3 \, {\cal H} (1-A)+\frac{1}{2}\left(-6 \,
    \psi +2 \, \nabla^2 E\right)' +\nabla^2 \left(v-B\right)\right]+O(2) \, .
\end{equation}
At this point, we have all the ingredients to compute $\zeta_\mu^{(1)}$
\be 
\begin{split}
\label{wl}
&\zeta_i^{(1)}= \partial_i \left[-{\cal H}(\bar\eta_0) \eta_0^{(1)}-{\cal H}\frac{\rho^{(1)}}{\bar{\rho}}
-\frac{1}{3}(3\psi-\nabla^2 E)|_{\left(\bar\eta_0,x^i\right)}^{(\eta,x^i)}+\frac{1}{3}\int_{\bar\eta_0}^{\eta} \nabla^2 \left(v-B\right)\,d\eta \right]\, ,\\
& \zeta_0=-{\cal H}(\bar\eta_0) \partial_\eta\eta_0 +O(2)= O(2) \, .
\end{split}
\ee
Using the $\zeta$ definition
\be
\label{zeta_i}
\begin{split}
\zeta_i& =\partial_i \left(\zeta+\frac{1}{3}\nabla^2 E+\frac{1}{3}\int_{\bar{\eta}_0}^\eta \nabla^2 \left(v-B\right) \,d\eta+\psi(\bar{\eta}_0,x^i)+{\cal H}(\bar\eta_0)\, \frac{f^{(1)}}{\bar f'}(\bar{\eta}_0,x^i)-\frac{1}{3}\nabla^2 E(\bar{\eta}_0,x^i) \right) \\
& = \partial_i \hat{\zeta}_s \, ,
\end{split}
\ee
thus, also in the case of a generic perfect fluid we find a scalar $\zeta_{s}^{(1)}$ such that $\zeta_i=\partial_i \zeta_s$.
At this point we can show that in the case of a perfect and barotropic fluid, the $\zeta_i$ time dependence is completely fictitious. Indeed,
from the standard relation
\be
\zeta'=-\frac{1}{3}\nabla^2 \left(E'+v-B\right)- \frac{{\cal H}}{\bar{\rho}+\bar{p}}\Gamma^{(1)} \, ,
\ee
integrating we get
\be 
\label{con}
\zeta+\frac{1}{3}\nabla^2 E+\frac{1}{3} \int_{\bar{\eta}_0}^{\eta} \nabla^2 \left(v-B\right)\,d\eta'=\zeta(\bar{\eta}_0,x^i)+\frac{1}{3}\nabla^2 E(\bar{\eta}_0,x^i) - \int_{\bar{\eta}_0}^{\eta} \frac{{\cal H}}{\bar{\rho}+\bar{p}}\Gamma^{(1)} \,d\eta' \, .
\ee
Substituting Eq. (\ref{con}) in Eq. (\ref{zeta_i}) we get
\be 
\label{fin_zeta}
\begin{split}
\hat{\zeta}_{s}^{(1)} &= {\cal H}(\eta_0)\, \frac{f^{(1)}}{\bar f'}(\bar{\eta}_0,x^i)+\psi(\bar{\eta}_0,x^i)+\zeta(\bar{\eta}_0,x^i)- \int_{\bar\eta_0}^{\eta} \frac{{\cal H}}{\bar{\rho}+\bar{p}}\Gamma^{(1)}\, d\eta' \\
&= \zeta_s^{(1)}-\int_{\bar\eta_0}^{\eta} \frac{{\cal H}}{\bar{\rho}+\bar{p}}\Gamma^{(1)}\, d\eta' 
\end{split}
\ee
In the case of a barotropic perfect fluid ($\Gamma^{(1)}=0$),
Eq. (\ref{fin_zeta}) coincides with Eq. (\ref{gen_f}), showing that
there is no time dependence and the perturbative approach is coherent
with our result (\ref{barres}).  

The same conclusion is reached proceeding as in ~\cite{Langlois:2005qp}; by expanding the definition $\theta =3 u^a \partial_a {\cal N}$, we
get 
\be
\theta=\frac{1}{a} \left[3{\cal H}(1-A)+3{\cal N}'^{(1)}\right] \, .
\ee
By comparison with Eq. (\ref{the}), one can check that ${\cal N}'^{(1)}$ has the following form
\be
\label{alpha'}
{\cal N}'^{(1)}=-\psi'+\frac{1}{3}\nabla^2 (E'+v-B) \, ;
\ee
thus 
\be
\begin{split}
\label{alpha}
{\cal N}^{(1)}&=\int\left(\frac{1}{3}\nabla^2 (E'+v-B)-\psi' \right)d\eta'+L(x^i)\\
            &=-\psi+\frac{1}{3}\nabla^2
            E+\frac{1}{3}\int_{\bar{\eta}_0}^\eta \nabla^2
            \left(v-B\right) 
\, d\eta'+\psi(\bar{\eta}_0,x^i)-\frac{1}{3}\nabla^2 E(\bar{\eta}_0,x^i)-{\cal H}(\eta_0)\eta_0^{(1)} \, .
\end{split}
\ee
In the second line of Eq. (\ref{alpha}), we set the arbitrary function
of the spatial coordinates $L$~\footnote{In the original paper ~\cite{Langlois:2005qp}, actually it was set $L=0$.} equal to 
$-{\cal H}(\bar{\eta}_0)\eta_0^{(1)}$. This time-independent function
is a first-order contribution coming from perturbing the {\it background} number of e-folds
$\ln\left(\frac{a(\eta)}{a(\eta_0)}\right)$. With such a choice we recover the expression (\ref{zeta_i}).

\subsection{2-nd order, barotropic fluid}
\label{sec_zeta}
In this Appendix the second-order $\zeta_\mu$ expression in the case of perfect and barotropic fluid is computed starting from our exact result (\ref{barres}), using initial constant-proper-time hyper-surfaces for which we have found second order contributions in Appendix \ref{x0}. 
Let us start by denoting with $g$ a generic first-order physical quantity, such that
\be 
 g|_0=g\left(\bar{\eta}_0+\delta \eta_0\, , x^j+\delta_x^j\right)=g(\bar{\eta}_0,x^i)+(\partial_\mu g) |_{\bar{0}}\, \delta^{\mu}\,, \qquad \delta^{\mu}= \left(\delta \eta_0,\,\delta_x^j\right)\, . 
\ee
Where $\delta\eta_0= \eta_0^{(1)}+\eta_0^{(2)}+O(3)$ and $\delta_x^j= x_0^j-x^j=-\int_{wl} \, d\tau\, u^j$. 
Using this simple Taylor expansion of quantities computed on the initial hyper-surface and simply remembering that we are dealing with composed functions
\be
\partial_\mu g \left(x_0 \left(x\right)\right)= \partial_\alpha g |_0 \,\partial_\mu x_0^\alpha \, ,
\ee 
we can analyse the second-order terms inside Eq. (\ref{barres}).
Indeed, perturbing up to the second order the $\partial_i \rho(x_0)$ term we find
\be
\begin{split}
\zeta_i &= \zeta_i^{(1)}+\zeta_i^{(2)}\\
        &= \zeta_i^{(1)}-\frac{1}{3(\bar\rho+\bar p)^2|_{\bar 0}}\left(p^{(1)}+\rho^{(1)}+v^{(1)} (\bar{\rho}+\bar p)'\right)\left(\partial_i\rho^{(1)}+\bar \rho ' \partial_i v^{(1)}\right)|_{\bar 0}\\
        &\;\;\;+\frac{1}{3(\bar \rho+ \bar p)|_{\bar 0}}\left[ \partial_i \left(\partial_j \rho^{(1)}\, \delta_x^{j\,(1)}\right)+\bar{\rho}' \partial_i \eta_0^{(2)}+\partial_i\rho^{(2)}+\partial_i \left(\rho'^{(1)}v^{(1)}\right)+\frac{1}{2}\bar{\rho}'' v^{(1)\, 2}\right]|_{\bar{0}} \, .
\end{split}
\ee
Using the fluid barotropicity $p^{(1)}=w(\rho)\rho^{(1)}$, and putting into evidence a spatial gradient, we get that in the case of a barotropic perfect fluid, up to the second perturbative 
order we can write $\zeta_i=\partial_i(\zeta_{s}^{(1)}+\zeta_{s}^{(2)})$, where
\be
\label{zetaVL_2}
\zeta_s^{(2)}= -\frac{1}{2}\left\lbrace 3(1+w)\zeta_s^{(1)\,2}+\frac{{\cal H}}{\bar\rho '}\left[\left(\rho^{(2)}+\rho'^{(1)}v^{(1)}+\frac{1}{2}\bar{\rho}'' v^{(1)\, 2}\right)|_{\bar{0}}+\bar{\rho}'|_{\bar{0}}\eta_0^{(2)}+\partial_j \rho^{(1)}|_{\bar{0}}\, \delta_x^{j\,(1)}\right]\right\rbrace \, 
\ee
and
\be
\delta_x^{j \, (1)}= -\int_{wl} \, d\eta' a(\eta') \,u^{j \, (1)} \, ,  
\ee
while $\eta_0^{(2)}$ is a function characterized by Eq. (\ref{eta_2}). The $\zeta_s^{(2)}$ expression is particularly simple at large scales, where neglecting terms with two spatial 
derivatives we find
\be
\zeta_s^{(2)}= \frac{{\cal H}}{\bar\rho '}\left[\left(\rho^{(2)}+\rho'^{(1)}v^{(1)}+\frac{1}{2}\bar{\rho}'' v^{(1)\, 2}\right)|_{\bar{0}}+\bar{\rho}'|_{\bar{0}}\eta_0^{(2)}\right] \, ,
\ee 
which has a very compact form in a gauge where $v=0$
\be
\zeta_s^{(2)}= \frac{{\cal H}}{\bar\rho '}\rho^{(2)}|_{\bar{0}} \, .
\ee 
The $\zeta_0$ computation is completely analogous, noting that:
\be 
\begin{split}
\partial_\eta (\rho)|_0 &= \partial_{\eta_0} \, \rho \partial_\eta \eta_0 +\partial_{j_0} \rho \, \partial_\eta x_0^j \\
&=\rho'|_{\bar{0}}\, \partial_\eta \eta_0^{(2)}+\left(\partial_{j} \rho^{(1)}\right) |_{\bar{0}} \,\partial_\eta x_0^j+O(3) \, ,
\end{split}
\ee
therefore
\be 
\zeta_0^{(2)}=-{\cal H}(\bar{\eta}_0) \partial_\eta \left( \eta_0^{(2)}+\frac{\partial_j\rho^{(1)}}{\bar{\rho}'}|_{\bar{0}} x_0^j\right) \, .
\ee
Using the $\partial_\eta \eta_0$ and $\partial_\eta x_0^j$ expressions obtained in Appendix \ref{x0} and substituting in $\zeta^{0\,(2)}$
\be
\begin{split}
\label{zeta_0}
\zeta_0^{(2)}&= -\frac{{\cal H}(\bar{\eta}_0)}{\bar{\rho}'(\bar{\eta}_0)}\partial_j\left[v\bar{\rho}'|_{\bar{0}}+\rho^{(1)}|_{\bar{0}}\right]\partial_\eta x_0^j\\
&= \zeta_j^{(1)}\partial_\eta x_0^j\\
& = \bar{u}_0 \,u^{j\, {(1)}}\zeta_j^{(1)} \, .
\end{split}
\ee
Notice that this relation holds also in the case of a generic perfect fluid. Indeed, starting from the identity
\be
D_0 {\cal N}= u_0\, u^i\partial_i {\cal N}-u_i u^i {\cal N}' \, ,
\ee 
we get
\be
\begin{split}
\zeta_0^{(2)}&= D_0 {\cal N}- \frac{\dot{{\cal N}}}{\dot{\rho}}D_0\rho \\ 
&=\bar{u}_0 u^{i\, (1)}\left[\partial_i {\cal N}^{(1)}-\frac{\bar{{\cal N}}'}{\bar{\rho}'}\partial_i \rho^{(1)}\right]\\
&=\bar{u}_0 \,u^i \zeta_i^{(1)} \, ,
\end{split}
\ee
which is the same result obtained in Eq. (\ref{zeta_0}).\\
Finally, as a further check, ket us show that our result (\ref{barres}) is consistent with ${\cal L}_u \, \zeta_\mu=0$, order-by-order in perturbation theory. At linear order we have
\be
\zeta_\mu^{(1)}{}' =0 \, .
\ee
We have
\be
\begin{split}
{\cal L}_u \zeta_\mu &= u^\nu \partial_\nu \zeta_\mu +\zeta_\nu\partial_\mu u^\nu \\
&=\bar{u}^0 \zeta_\mu'\,^{(1)}+\bar{u}^0
\zeta_\mu'\,^{(2)}+u^{0\,(1)} \zeta_\mu'\,^{(1)}+u^{i \,
  (1)} \partial_i \zeta_\mu^{(1)}+\zeta_0^{(2)}\partial_\mu
\bar{u}^0+\zeta_i^{(1)}\partial_\mu u^{i\,(1)}
+O(3)\\
&=\bar{u}^0 \zeta_\mu'\,^{(2)}+u^{i \, (1)} \partial_i
\zeta_\mu^{(1)}+\zeta_0^{(2)}\partial_\mu
\bar{u}^0+\zeta_i^{(1)}\partial_\mu u^{i\,(1)} +O(3)=0 \, .
\end{split}
\ee
Therefore
\be
\label{Lie}
\bar{u}^0 \zeta_i'\,^{(2)}=-u^{j\,(1)}\partial_j\zeta_i^{(1)}-\zeta_j^{(1)}\partial_i u^{j \, (1)}\, , 
\ee
\be
\bar{u}^0 \zeta_0'\,^{(2)}=-\bar{u}'\,^{0} \zeta_0^{(2)}-u'\,^{i\,(1)} \zeta_i^{(1)}\, .
\ee
As a matter of fact, (\ref{zetaVL_2}) and (\ref{zeta_0}) satisfy the above relations.

\bibliographystyle{unsrt}  
\bibliography{paper}

\end{document}